\documentclass[twocolumn,showpacs,preprintnumbers,amsmath,amssymb,floatfix]{revtex4}
\usepackage{graphicx}

\usepackage{bm}
\usepackage{dcolumn}

\begin{document}

\title{Spreading Dynamics of Polymer Nanodroplets}

\author{David R. Heine, Gary S. Grest, and Edmund B. Webb III}

\affiliation{Sandia National Laboratories, Albuquerque, New Mexico 87185}

\date{\today}

\begin{abstract}
The spreading of polymer droplets is studied using molecular dynamics
simulations. To study the dynamics of both the precursor foot and
the bulk droplet, large drops of $~200,000$ monomers are simulated
using a bead-spring model for polymers of chain length $10,$ $20,$
and $40$ monomers per chain. We compare spreading on flat and atomistic
surfaces, chain length effects, and different applications of the
Langevin and dissipative particle dynamics thermostats. We find diffusive
behavior for the precursor foot and good agreement with the molecular
kinetic model of droplet spreading using both flat and atomistic surfaces.
Despite the large system size and long simulation time relative to
previous simulations, we find no evidence of hydrodynamic behavior
in the spreading droplet.
\end{abstract}

\maketitle

\section{Introduction}

The spreading of liquid droplets on a surface is an important issue
for several industries including adhesion, lubrication, coating, and
printing. Emerging nanotechnology in areas such as lithography and
microfluidics has made the issue of droplet spreading on small length
scales even more relevant. Experiments on droplet spreading have revealed
several phenomena involved in the spreading process, some of which
occur on the atomic level and others that become relevant at mesoscopic
length scales \cite{HS:JCI:71,T:JPD:79,G:RMP:85,HFC:Nat:89,RCB:Lan:97,VVO:Lan:98,GCS:Lan:98,RCO:Lan:99,SBG:Lan:99,RCV:Lan:00,VC:AM:00,PSS:JCI:01,BS:JCI:02,CBC:Lan:02,CBP:Lan:02}.
These include the spreading of a precursor foot ahead of the droplet
\cite{G:RMP:85}, terraced spreading of mono-molecular layers \cite{HFC:Nat:89,HCL:PRL:90,FVC:JCI:93},
and viscous losses due to rolling motion \cite{HS:JCI:71,DD:JFM:74}.

Several models have been proposed to describe the spontaneous spreading
of liquid droplets on a surface. These models can be classified as
molecular kinetic models, continuum hydrodynamic models, or combined
models. The molecular kinetic theory of Eyring \cite{GLE:TRP:41}
has been applied to the kinetics of wetting by Blake and Haynes \cite{B:CAT:68,BH:JCI:69}
as well as Cherry and Holmes \cite{CH:JCI:69}. This theory treats
the surface adsorption of liquid molecules as the dominant factor
in the spreading of a droplet. The hydrodynamic theory \cite{HS:JCI:71,V:FD:76,T:JPD:79,C:JFM:86}
focuses on the energy dissipation due to viscous flow in the droplet.
It has been claimed that hydrodynamic dissipation is dominant for
small contact angles and non-hydrodynamic dissipation is dominant
for relatively large contact angles \cite{BG:ACI:92}. Since both
mechanisms are present in spreading droplets, several groups have
proposed combined theories \cite{V:FD:76,G:RMP:85,G:LI:90,PP:Lan:92,B:Wet:93,RCO:Lan:99}.
Experimental results for the spreading of poly(dimethylsiloxane) (PDMS)
drops on bare silicon wafers have shown good agreement with one combined
model \cite{RCV:Lan:00}.

The study of droplet spreading using molecular dynamics simulation
has been hindered due to computational limitations restricting simulations
to small droplet sizes and short times. Molecular dynamics simulations
were first used to study the spreading of monomer and dimer liquids
\cite{TGW:JCP:84,YKB:PRL:91,YKB:PRA:92,RMS:JCP:99}. However, the
spreading of monomer and dimer droplets are clearly influenced by
the volatility of the small molecules, allowing them to vaporize and
condense independent of the dynamics of the droplet. To separate the
spreading from the vaporization and condensation, subsequent simulations
used short bead-spring chain molecules since they have a very low
vapor pressure. In most cases, the simulations reproduced the experimentally
observed $R\sim t^{1/2}$ scaling of the contact radius of the precursor
foot on both atomistic \cite{COK:PRL:95,OCK:PRE:96,VC:AM:00,WPF:PF:03}
and flat \cite{NAK:PRL:92,NA:PRE:94} surfaces, though logarithmic
scaling has also been observed \cite{WPF:PF:03}. It is believed that
this difference is due to the corrugation of the substrate, producing
$t^{1/2}$ scaling for a sufficiently small lattice dimension and
a logarithmic scaling for large, i.e. strong corrugation \cite{BKV:PRL:96}.
Milchev and Binder \cite{MB:JCP:02} have studied wetting using Monte
Carlo simulations on a flat substrate which suggest Tanner's spreading
law for the growth dynamics of the droplet holds on the nanoscopic
scale. Other comparisons to theoretical models have strongly supported
the molecular kinetic theory of wetting \cite{HCL:PRL:89,BCC:Lan:97,RBC:Lan:99,BCC:CSA:99,RBC:JPS:99},
probably due to the relatively small droplet sizes and short simulation
times employed.

In this paper, we present results from extensive molecular dynamics
simulations of coarse-grained models of polymer droplets wetting a
surface. Although most recent simulations of droplet spreading use
droplets containing $20,000$ to $32,000$ monomers \cite{COK:PRL:95,OCK:PRE:96,BCC:Lan:97,BCC:CSA:99,VC:AM:00,WPF:PF:03},
we consider drops composed of $100,000$ to $200,000$ monomers to
simultaneously study the precursor foot and bulk regions for long
times. We compare simulations performed using both a flat surface
and an atomistic substrate to determine if the computationally expensive
atomistic substrate is required to obtain correct spreading dynamics.
We also evaluate different implementations of the Langevin and dissipative
particle dynamics (DPD) thermostats for efficiency and realism in
preserving hydrodynamic effects. Also, the difference in using a spherical
droplet as the starting configuration as opposed to a hemispherical
droplet is discussed. We find that the method which captures all of
the physics of the spreading drop in the most computationally efficient
manner is to simulate large drops on flat substrates with a coupling
to the thermostat which falls off exponentially with distance from
the substrate \cite{BP:PRE:01}. For atomistic substrates, we find
coupling only the substrate monomers to the Langevin thermostat significantly
more efficient than coupling the DPD thermostat to all monomers.

The paper is organized as follows. Section \ref{sec:Simulation-Details}
describes the details of the molecular dynamics simulations and the
application of the thermostats. Section \ref{sec:Simulation-Results}
presents the results for the time dependence of the contact radius.
The contact angle data is fit to models of droplet spreading in Section
\ref{sec:Models-of-Droplet} and conclusions are presented in Section
\ref{sec:Conclusions}.

\section{\label{sec:Simulation-Details}Simulation Details}

\subsection{System}

We perform molecular dynamics (MD) simulations using a coarse-grained
model for the polymer chains in which the polymer is represented by
spherical beads of mass $m$ attached by springs. We use a cutoff
Lennard-Jones (LJ) potential to describe the interaction between all
monomers. The LJ potential is given by

\begin{equation}
U_{LJ}^{\alpha\beta}(R)=\left\{ \begin{array}{rl}
4\varepsilon_{\alpha\beta}\left[\left(\frac{\sigma_{\alpha\beta}}{r}\right)^{12}-\left(\frac{\sigma_{\alpha\beta}}{r}\right)^{6}\right] & r\leq r_{c}\\
0 & r>r_{c}\end{array}\right.\label{eq:ljcut}\end{equation}
 where $\varepsilon_{\alpha\beta}$ and $\sigma_{\alpha\beta}$ are
the LJ units of energy and length and the cutoff is set to $r_{c}=2.5\:\sigma_{\alpha\beta}$.
We denote the polymer monomers as type 1 and substrate monomers as
type 2. The monomer-monomer interaction, $\varepsilon_{11}=\varepsilon$,
is used as the reference and all monomers have the same diameter $\sigma_{\alpha\beta}=\sigma$.
For bonded monomers, we apply an additional potential where each bond
is described by the finite extensible nonlinear elastic (FENE) potential
\cite{KG:JCP:90},\begin{equation}
U_{FENE}(r)=\left\{ \begin{array}{rl}
\frac{-k}{2}R_{0}^{2}\ln\left[1-\left(\frac{r}{R_{0}}\right)^{2}\right] & r\leq R_{0}\\
\infty & r>R_{0}\end{array}\right.,\label{eq:fene}\end{equation}
 with $k=30\:\varepsilon$ and $R_{0}=1.5\:\sigma$.

Droplets consisting of chains of length $N=10,$ $20,$ or $40$ monomers
per chain are created by first equilibrating a melt of the polymer
and then removing molecules whose centers are outside of a hemisphere
of a given radius, $38\:\sigma$ for non-wetting droplets and $48\:\sigma$
for wetting droplets. The droplet is then placed on either an atomistic
substrate or a flat substrate.

The atomistic substrate is composed of LJ particles forming four layers
of the $(111)$ surface of an fcc lattice where the bottom layer is
frozen and the top three layers maintain their structure through a
strong LJ interaction, $\varepsilon_{22}=5\varepsilon$. The masses
of the substrate monomers are set to $m_{2}=2m_{1}=2m$. For non-wetting
droplets, each layer of the substrate contains $12\,000$ monomers
and the dimensions of the substrate are $110.0\sigma\:\times\:115.4\sigma$.
For the wetting droplets, we study two substrates, containing either
$49\,200$ or $99\,960$ monomers per layer. The dimensions of the
substrates are $231.2\:\sigma\:\times\:231.0\:\sigma$ and $330.8\:\sigma\:\times\:331.4\:\sigma$,
respectively. We refer to these as the small, medium and large substrates.
The large substrates are necessary because the finite size of the
atomistic substrates require the use of periodic boundary conditions
at their edges whereas the flat surface can extend indefinitely in
the $x$ and $y$ directions. For the atomistic substrate, during
the course of the simulation, the precursor foot reaches the edge
of the substrate and interacts with the periodic image of the droplet.
Although this can be related to the spreading of an array of nanodroplets,
such as in micro-contact printing, we do not include any data for
the precursor foot once it reaches the periodic image. The droplets
consist of $\sim100,000$ monomers for non-wetting droplets and $\sim200,000$
monomers for wetting droplets. All simulations are run at a temperature
of $T=1.0\:\varepsilon/k_{B}$.

For the flat surface, the interaction between the monomers in the
droplet and the surface is modeled by an integrated LJ potential,

\begin{equation}
U_{LJ}^{wall}(z)=\left\{ \begin{array}{rl}
\frac{2\pi\varepsilon_{w}}{3}\left[\frac{2}{15}\left(\frac{\sigma}{z}\right)^{9}-\left(\frac{\sigma}{z}\right)^{3}\right] & z\leq z_{c}\\
0 & z>z_{c}\end{array}\right.\label{eq:ljwall}\end{equation}
 with $z_{c}=2.2\sigma.$

The equations of motion are integrated using a velocity-Verlet algorithm.
We use a time step of $\Delta t=0.009\;\tau$ where $\tau=\sigma\left(\frac{m}{\varepsilon}\right)^{1/2}$.
The simulations are performed using the \textsc{lammps} code \cite{P:JCP:95}
on 36 to 100 Dec Alpha processors of Sandia's $CPlant^{®}$ cluster.
Simulating one million steps for a wetting drop of $200,000$ monomers
on the medium atomistic substrate takes between $90$ and $250$ hours
on $64$ processors, depending on the thermostat.

\subsection{Thermostats}

The choice of thermostat employed can greatly affect the droplet spreading
dynamics, so we compare simulations that use the Langevin \cite{GK:PRA:86}
and DPD \cite{HK:EL:92,EW:EL:95} thermostats. The purpose is to find
an approach that is both computationally efficient and provides a
realistic representation of the transfer of energy in the spreading
droplet.

The Langevin thermostat simulates a heat bath by adding Gaussian white
noise and friction terms to the equation of motion,

\begin{equation}
m_{i}\mathbf{\ddot{r}}_{i}=-\Delta U_{i}-m_{i}\gamma_{L}\dot{\mathbf{r}_{i}}+\mathbf{W}_{i}(t),\label{eq:lang}\end{equation}
 where $\gamma_{L}$ is the friction parameter for the Langevin thermostat,
$-\Delta U_{i}$ is the force acting on monomer $i$ due to the potentials
defined above, and $\mathbf{W}_{i}(t)$ is a Gaussian white noise
term such that

\begin{equation}
\left\langle \mathbf{W}_{i}(t)\cdot\mathbf{W}_{j}(t')\right\rangle =6k_{B}Tm_{i}\gamma_{L}\delta_{ij}\delta\left(t-t'\right).\label{eq:whitenoise}\end{equation}
The Langevin thermostat can either be coupled to all monomers in the
system or just to those in the substrate. The advantage of the latter
is that the long-range hydrodynamic interactions are preserved in
the droplet, whereas coupling all monomers to the Langevin thermostat
screens the hydrodynamic interactions. Both approaches are applied
in the simulations to test the various models for droplet spreading
discussed below in Sec.~IV. The damping constant is chosen to be
$\gamma_{L}=0.1\:\tau^{-1}$ in most cases, which is much smaller
than that arising from collisions between monomers.

Our next approach is to apply the thermostat from the DPD simulation
method. The DPD technique includes a dissipative force term in the
equations of motion along with random forces. The equation of motion
for the DPD thermostat is

\begin{equation}
m_{i}\mathbf{\ddot{r}}_{i}=\sum_{j\neq i}\left(-\Delta U_{ij}+\mathbf{F}_{ij}^{D}+\mathbf{F}_{ij}^{R}\right).\label{eq:dpd}\end{equation}
 In Eq. \ref{eq:dpd}, $\mathbf{F}_{ij}^{D}$ and $\mathbf{F}_{ij}^{R}$
are the dissipative and random terms given by

\begin{equation}
\mathbf{F}_{ij}^{D}=-m_{i}\gamma_{DPD}w^{2}(r_{ij})\left(\hat{\mathbf{r}}_{ij}\cdot\left(\dot{\mathbf{r}}_{i}-\dot{\mathbf{r}}_{j}\right)\right)\hat{\mathbf{r}}_{ij}\label{eq:diss}\end{equation}
 \begin{equation}
\mathbf{F}_{ij}^{R}=m_{i}\sigma_{DPD}w(r_{ij})\zeta_{ij}\hat{\mathbf{r}}_{ij}\label{eq:rand}\end{equation}
where $\gamma_{DPD}$ is the DPD friction parameter, $\sigma_{DPD}^{2}=2k_{B}T\gamma_{DPD}$,
$\zeta_{ij}$ is a Gaussian noise term with $\left\langle \zeta_{ij}(t)\zeta_{kl}(t')\right\rangle =\left(\delta_{ik}\delta_{jl}+\delta_{il}\delta_{jk}\right)\delta\left(t-t'\right)$,
$\mathbf{r}_{ij}=\mathbf{r}_{i}-\mathbf{r}_{j}$, $r_{ij}=\left|\mathbf{r}_{ij}\right|$,
and $\hat{\mathbf{r}}_{ij}=\mathbf{r}_{ij}/r_{ij}$. The weight function
$w(r_{ij})$ is defined as

\begin{equation}
w(r_{ij})=\left\{ \begin{array}{cl}
\left(1-r_{ij}/r'_{c}\right) & r_{ij}<r'_{c}\\
0 & r_{ij}\geq r'_{c}.\end{array}\right.\label{eq:weightfcn}\end{equation}
 We take $r'_{c}=r_{c}=2.5\:\sigma$. The advantage of this thermostatting
technique is that the momentum is conserved locally and long-range
hydrodynamic interactions are preserved even in the case where all
monomers are coupled to the thermostat. All simulations with the DPD
thermostat use $\gamma_{DPD}=0.1\:\tau^{-1}$, so the dissipation
from the thermostat is much less than from monomer collisions as seen
in Sec. \ref{sec:Models-of-Droplet}. Simulations that use DPD couple
the thermostat to all atoms in the system.

In the case of the flat substrate, we study several methods to thermostat
the system. In the first, we simply couple the Langevin or DPD thermostat
to all monomers. However this is somewhat unphysical since monomers
near the substrate are expected to have a stronger damping than those
in the bulk of the droplet. In the case of the Langevin thermostat,
this coupling of all monomers also means that the hydrodynamic interactions
are screened. In addition, chains which separate from the droplet
move across the substrate very rapidly, particularly for the DPD thermostat.
For this reason, we did not further pursue the DPD thermostat on the
flat substrate. To overcome these difficulties, we follow the approach
of Braun and Peynard \cite{BP:PRE:01} and add an external Langevin
coupling with a damping rate that decreases exponentially away from
the substrate. We choose the form

\begin{equation}
\gamma_{L}(z)=\gamma_{L}^{s}\exp\left(\sigma-z\right)\label{eq:expdamp}\end{equation}
 where $\gamma_{L}^{s}$ is the surface Langevin coupling and $z$
is the distance from the substrate. We choose values of $\gamma_{L}^{s}=1.0$,
$3.0$, and $10.0\:\tau^{-1}$. There is no obvious a priori way to
define the appropriate value of $\gamma_{L}^{s}$. However, one way
is to choose $\gamma_{L}^{s}$ so that the diffusion constant of the
precursor foot is comparable for the flat and atomic substrates for
comparable departures from the wetting/non-wetting transition (see
Fig. \ref{cap:modelfit} below).

\section{\label{sec:Simulation-Results}Simulation Results}

\begin{figure}

\includegraphics[width=8.5cm]{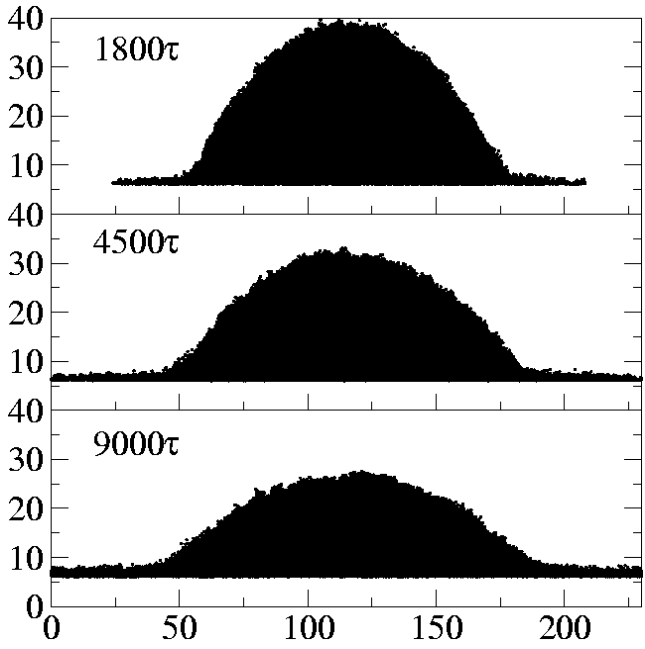}
\caption{\label{cap:configsub}Profile of the $N=10$ polymer droplet spreading
on the atomistic substrate at three different times using the Langevin
thermostat applied only to the substrate monomers with $\gamma_{L}=0.1\:\tau^{-1}$
and $\varepsilon_{12}=1.5\:\varepsilon$.}

\end{figure}

\begin{figure}

\includegraphics[width=8.5cm]{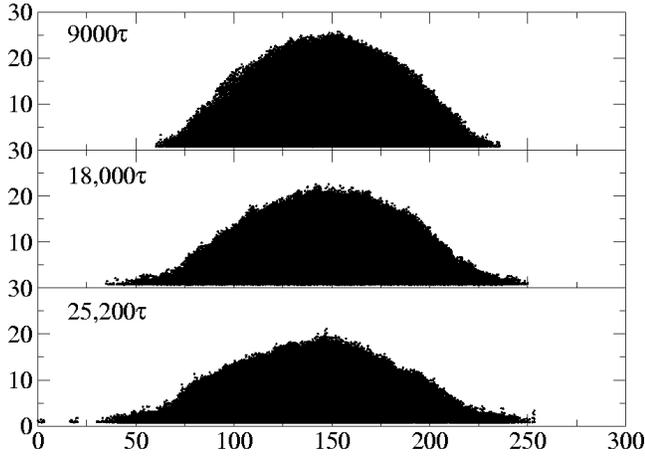}
\caption{\label{cap:configflat}Profile of the $N=10$ polymer droplet spreading
on the flat surface at three different times using the surface Langevin
thermostat. $\gamma_{L}^{s}=10.0\:\tau^{-1}$, $\varepsilon_{w}=2.0\:\varepsilon$.}

\end{figure}

A droplet containing about $200,000$ monomers for a wetting droplet
is large enough to allow us to simultaneously study the bulk and precursor
foot regions. This can be seen in the profile views for chain length
$N=10$ in Figs. \ref{cap:configsub} and \ref{cap:configflat}, which
show the foot extending beyond the bulk region for wetting droplets
on an atomistic substrate and a flat surface, respectively. Note that
Fig. \ref{cap:configsub} shows the thickness of the foot increasing
after it reaches the periodic image. The same behavior is seen when
periodic boundaries are applied to the flat surface, so this is not
an effect of the corrugation of the substrate.

To characterize the spreading dynamics of these droplets, we extract
the instantaneous contact radius and contact angle every $10,000$
to $40,000$ $\Delta t$. The contact radius is calculated by defining
a two-dimensional radial distribution function, $g(r)=\rho(r)/\rho$,
based on every particle within $1.5\:\sigma$ of the surface. The
local density at a distance $r$ from the center of mass of the droplet
is\begin{equation}
\rho(r)=\frac{N(r)}{2\pi r\Delta r}\label{eq:rdf}\end{equation}
 where $N(r)$ is the number of particles at a distance between $r$
and $r+\Delta r$ from the center of mass and $\rho$ is the integral
of $\rho(r)$ over the entire surface. The contact radius is defined
as the distance $r$ at which $g(r)=0.98$. This approach provides
a robust measure of the radius at any point during the spreading simulation.
The same calculation is used to obtain the droplet radius for ten
slices of the droplet at incremental heights every $1.5\:\sigma$
from the surface. A line is fit to the resulting points and the instantaneous
contact angle is determined from the slope of the line. For simulations
that exhibit a precursor foot, the particles within $4.5\:\sigma$
of the surface are ignored in the contact angle calculation.

\begin{figure}

\includegraphics[width=8.5cm]{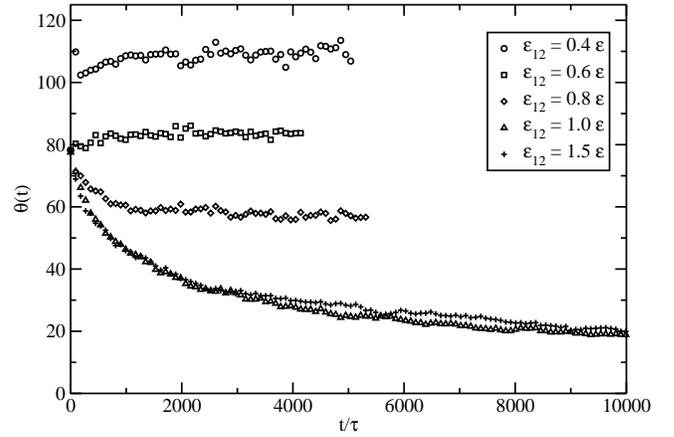}
\caption{\label{cap:nwflatcont}Contact angle of non-wetting droplets of $N=10$
polymers on an atomistic substrate starting from a hemispherical droplet
with $\gamma_{DPD}=0.1\:\tau^{-1}$.}

\end{figure}

\begin{figure}

\includegraphics[width=8.5cm]{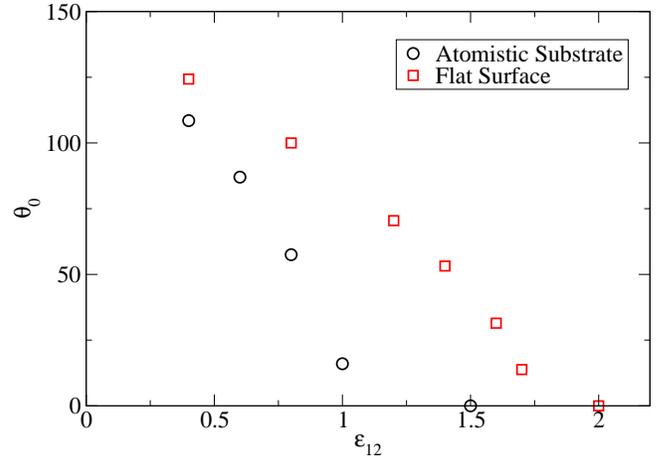}
\caption{\label{cap:modelfit}Equilibrium contact angle as a function of polymer-surface
interaction strength showing the transition from non-wetting to wetting
for $N=10$ polymer droplets on an atomistic substrate (circles) and
a flat surface (squares).}

\end{figure}

The non-wetting droplets reach their equilibrium configurations fairly
rapidly, as shown by the contact angle data in Fig. \ref{cap:nwflatcont}.
The equilibrium contact angle measured as a function of polymer-surface
interaction strength is shown in Fig. \ref{cap:modelfit}. From this
figure, it is clear that the transition from non-wetting to wetting
occurs near $\varepsilon_{12}^{c}\simeq1.05\:\varepsilon$ for droplets
on a substrate and $\varepsilon_{w}^{c}\simeq1.75\:\varepsilon$ for
droplets on a flat surface. For most of the wetting simulations, we
use $\varepsilon_{12}=1.5\:\varepsilon$ for the atomistic substrate
and $\varepsilon_{w}=2.0\:\varepsilon$ for the flat substrate, both
well within the wetting regime.

\begin{figure}

\includegraphics[width=8.5cm]{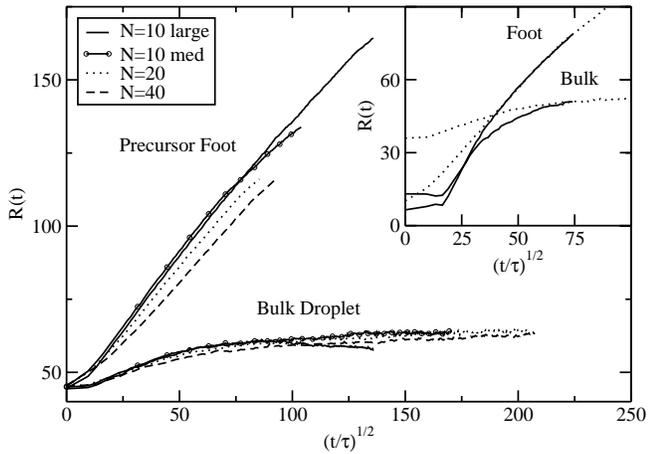}
\caption{\label{cap:radNsub}Time dependence of the contact radius of the
precursor foot and bulk droplet for wetting droplets on an atomistic
substrate at three different chain lengths starting from a hemisphere
with a contact angle of approximately $90^{o}$. The Langevin thermostat,
$\gamma_{L}=0.1\:\tau^{-1}$, is applied only to the substrate monomers
and $\varepsilon_{12}=1.5\:\varepsilon$. Results for $N=10$ are
for both the medium and large atomistic substrates, while those for
$N=20$ and $40$ are for the medium atomistic substrate. The inset
shows the contact radius for $N=10$ starting with a spherical droplet
(solid line) compared to a hemispherical droplet (dotted). Results
for the hemisphere in the inset have been shifted downward to easily
compare the late time behavior.}

\end{figure}

The time dependence of the contact radius of the precursor foot and
bulk region is shown in Fig. \ref{cap:radNsub} for wetting droplets
on an atomistic substrate for three chain lengths. The $t^{1/2}$
behavior is evident for the precursor foot at all chain sizes, while
the kinetics of the main droplet is clearly significantly slower.
The $N=10$ data shown in Fig. \ref{cap:radNsub} is taken from simulations
on both the large and medium substrates whereas the $N=20$ and $N=40$
simulations are on the medium substrate. The contact radius of the
bulk droplet increases steadily for all three chain lengths on the
medium substrate. However, the run on the large substrate shows a
slowing down and eventual contraction of the bulk contact radius as
the foot continues outward, depleting the supply of material in the
bulk faster than the drop can transfer material downward. This suggests
that for our largest substrate, the drop size must be even larger
to be able to study both the precursor foot and bulk droplet in the
same simulation.

The inset in Fig. \ref{cap:radNsub} shows the spreading of a spherical
$N=10$ droplet  compared to an initial hemisphere. The sphere is
placed just above the substrate with zero initial velocity to avoid
any effect due to impact velocity. The difficulty in measuring the
spreading rate for this case is evident as it takes roughly $1200\:\tau$
for the sphere to adopt a hemispherical shape, $1600\:\tau$ for the
spreading rate of the foot to match that of the hemisphere, and $5000\:\tau$
for the spreading rate of the bulk to match that of the hemisphere.
(The hemisphere data for the foot and bulk regions are shifted downward
to easily compare the spreading rates.) 

Vou\'{e} \textit{et al.} \cite{VVO:Lan:98,VC:AM:00} found both experimentally
for PDMS droplets and in numerical computer simulations that the diffusion
constant of the precursor foot varies non-monotonically with increasing
coupling to the substrate. At first, increasing the coupling to the
substrate increases the driving force and the fluid spreads on the
substrate more rapidly. However, further increases in the strength
of the fluid substrate coupling, while increasing the driving force,
also increase the friction of the fluid monomers with the substrate,
resulting in a decrease in the diffusion constant. From the time dependence
of $R(t)$, the diffusion constant $D_{f}$ for the foot can be determined
from\begin{equation}
\left\langle \left(R(t)-R(0)\right)^{2}\right\rangle =4D_{f}t\label{eq:diffconst}\end{equation}
 The resulting diffusion constants are $D_{f}=0.34\:\sigma^{2}/\tau$,
$0.30\:\sigma^{2}/\tau$, and $0.23\:\sigma^{2}/\tau$ for $N=10$,
$20$, and $40$ respectively for $\varepsilon_{12}=1.5\:\varepsilon$,
indicating a very weak dependence on chain length, at least for these
unentangled chains. Increasing $\varepsilon_{12}$ to $2.0\:\varepsilon$,
we find $D_{f}=0.16\:\sigma^{2}/\tau$ for $N=10$, thus the droplets
are in the high friction regime for these values of fluid substrate
coupling.

\begin{figure}

\includegraphics[width=8.5cm]{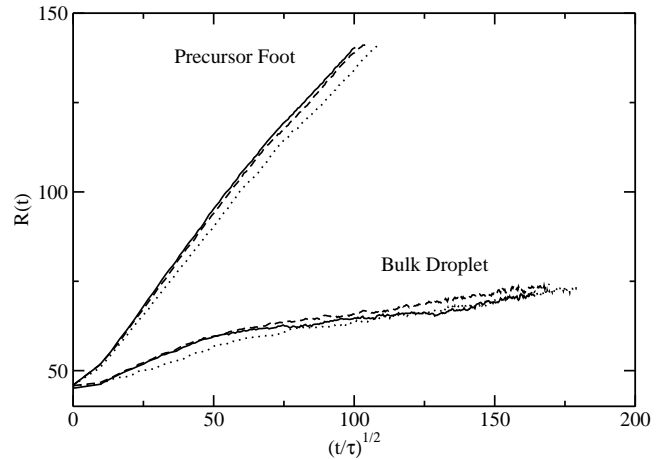}
\caption{\label{cap:radthermsub}Effect of thermostat on contact radius of
precursor foot and bulk region for wetting droplets of $N=10$ polymers
on an atomistic substrate for $\varepsilon_{12}=1.5\:\varepsilon$.
The thermostats applied are DPD (solid line), Langevin on all monomers
(dotted) and Langevin on only substrate monomers (dashed). $\gamma_{DPD}=\gamma_{L}=0.1\:\tau^{-1}.$}

\end{figure}

Figure \ref{cap:radthermsub} shows the time dependence of the contact
radius for wetting droplets on an atomistic substrate using different
thermostatting techniques. These results show that there is essentially
no difference in the spreading rate between the DPD thermostat applied
to all monomers and the Langevin thermostat applied only to the substrate.
We can see that applying the Langevin thermostat to all monomers slightly
decreases the spreading rate as the viscous heating is removed from
the system, though the resulting loss of hydrodynamic flow, at least
for the droplet size studied here, has no significant impact.

\begin{figure}

\includegraphics[width=8.5cm]{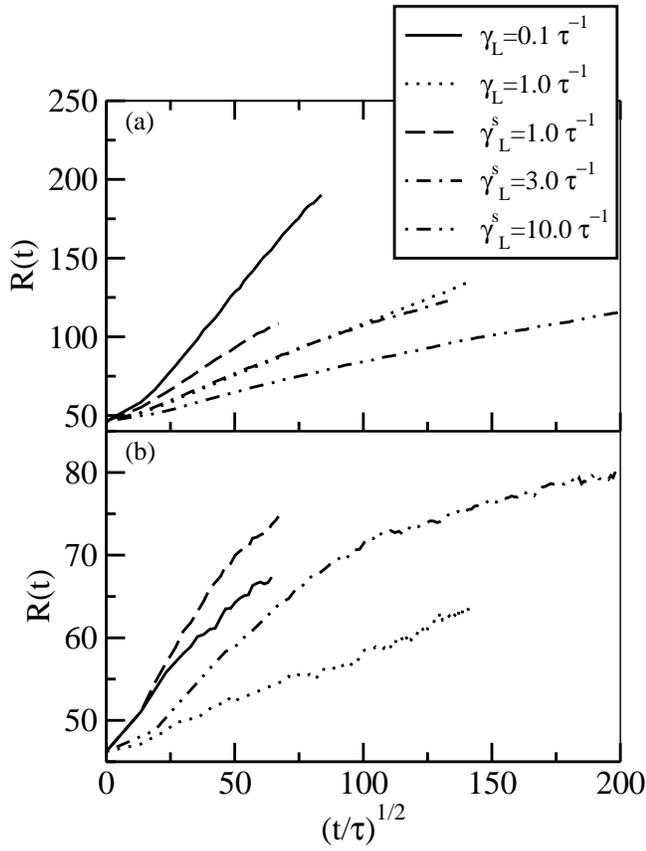}
\caption{\label{cap:radthermflat} Effect of thermostat on contact radius
of (a) precursor foot and (b) bulk region for wetting droplets of
$N=10$ polymers on a flat surface with $\varepsilon_{w}=2.0\:\varepsilon$. }

\end{figure}

\begin{figure}

\includegraphics[width=8.5cm]{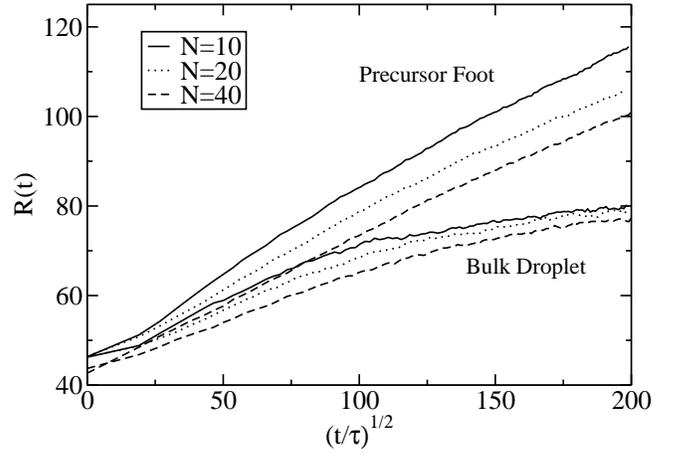}
\caption{\label{cap:radNflat}Chain length dependence of the contact radius
of the precursor foot and bulk droplet for wetting droplets on a flat
surface with $\varepsilon_{w}=2.0\:\varepsilon$. The surface Langevin
thermostat is applied with $\gamma_{L}^{s}=10.0\:\tau^{-1}$.}

\end{figure}

For wetting droplets on a flat surface, the thermostat dependence
of the contact radius is shown in Fig. \ref{cap:radthermflat}. Here,
the Langevin thermostat is applied either to all monomers (curves
labeled with $\gamma_{L}$) or with the surface Langevin coupling
(curves labeled with $\gamma_{L}^{s}$). The value of $\gamma_{L}$
clearly has a strong influence on the spreading rate. $\gamma_{L}^{s}=3.0\:\tau^{-1}$
gives a diffusion constant comparable to the atomistic substrate with
$\gamma_{L}=0.1\:\tau^{-1}$. The chain length dependence of the contact
radius is shown in Fig. \ref{cap:radNflat}. Again, the $t^{1/2}$
behavior is evident in the foot region but not the bulk region. The
chain length dependence on the flat surface is similar to the atomistic
substrate, showing a moderate decrease in spreading rate for larger
polymers.

\section{\label{sec:Models-of-Droplet}Models of Droplet Spreading Dynamics}

\subsection{Overview of models}

The dynamics of droplet spreading are controlled by the driving force
(the difference in surface tension $\gamma$ at each interface) and
by the energy dissipation. The total energy dissipation can be represented
by a sum of three different components, $T\left(\dot{\Sigma}_{w}+\dot{\Sigma}_{f}+\dot{\Sigma}_{l}\right)$
\cite{G:RMP:85}. The first term, $T\dot{\Sigma}_{w}$, represents
energy dissipation due to the hydrodynamic flow in the bulk of the
droplet as more material is transferred to the surface. $T\dot{\Sigma}_{f}$
relates to the viscous dissipation in the precursor foot present in
cases of complete wetting. The third term, $T\dot{\Sigma}_{l}$, refers
to the dissipation in the vicinity of the contact line due to the
adsorption and desorption of liquid molecules to the solid surface.
Here, we compare models that incorporate one or more of these dissipation
mechanisms to our simulation results.

The molecular kinetic theory of liquids developed by Eyring and coworkers
\cite{GLE:TRP:41} has been applied to droplet spreading by Blake
and Haynes \cite{BH:JCI:69}. It focuses on the adsorption of liquid
molecules to the surface as the dominant factor in energy dissipation.
In this theory, the liquid molecules jump between surface sites separated
by a distance $\lambda$ with a frequency $K$. The velocity of the
contact line is related to the contact angle $\theta$ by \begin{equation}
\frac{dR}{dt}=2K\lambda\sinh\left[\left(\frac{\gamma}{2\Delta nk_{B}T}\right)\left(\cos\theta-\cos\theta_{0}\right)\right]\label{eq:Blakev}\end{equation}
 where $\gamma$ is the surface tension of the liquid/vapor interface,
$\Delta n$ is the density of sites on the solid surface, and $\theta_{0}$
is the equilibrium contact angle. For sufficiently low velocities,
the equation can be written in its linearized form,\begin{equation}
\frac{dR}{dt}=\frac{K\lambda\gamma}{\Delta nk_{B}T}\left(\cos\theta_{0}-\cos\theta\right).\label{eq:Blakevlin}\end{equation}
 Assuming the droplet maintains constant volume and the shape of a
spherical cap, the velocity of the contact line can be expressed in
terms of the time dependence of the contact angle purely from geometric
arguments giving\begin{equation}
\frac{dR}{dt}=-\left(\frac{3V}{\pi}\right)^{1/3}\frac{\left(1-\cos\theta\right)^{2}}{\left(2-3\cos\theta+\cos^{3}\theta\right)^{4/3}}\frac{d\theta}{dt}.\label{eq:drdt}\end{equation}
 Combining Eqs. \ref{eq:Blakevlin} and \ref{eq:drdt} gives an expression
for the time dependence of the contact angle,\begin{equation}
\frac{d\theta}{dt}=-\left(\frac{\pi}{3V}\right)^{1/3}\Omega\left(\theta\right)\frac{\gamma}{\zeta_{0}}\left(\cos\theta_{0}-\cos\theta\right)\label{eq:kindqdt}\end{equation}
 where\begin{equation}
\Omega\left(\theta\right)=\frac{\left(2-3\cos\theta+\cos^{3}\theta\right)^{4/3}}{\left(1-\cos\theta\right)^{2}}\label{eq:omega}\end{equation}
 and $\zeta_{0}$ is the friction coefficient defined as $\zeta_{0}=\frac{\Delta nk_{B}T}{K\lambda}$
, which has units of viscosity.

The hydrodynamic model \cite{C:JFM:86} describes the flow pattern
that forms in the bulk of the droplet as material is transferred to
the advancing contact line. This model can be obtained by solving
the equations of motion and continuity for the droplet described as
a cylindrical disk \cite{SB:JAP:94} instead of a spherical cap. Neglecting
the flow perpendicular to the surface and balancing the radial shear
stress at the top of the cylinder with the effective radial surface
tension, the velocity of the contact line is written as\begin{equation}
\frac{dR}{dt}=\frac{4\gamma V^{3}}{\pi^{3}\eta R^{9}}-\frac{\gamma\beta V}{2\pi\eta R^{3}}\label{eq:hydrodrdt}\end{equation}
 where $V$ is the droplet volume, $\eta$ is the viscosity of the
liquid, and $\beta=1-\cos\theta_{0}$. Equation \ref{eq:hydrodrdt}
is in agreement with Tanner's spreading law \cite{T:JPD:79} for completely
wetting systems $\left(\theta_{0}=0\right)$ and for non-wetting systems
with small equilibrium contact angles, giving $R\sim t^{1/10}$ at
long times. Instead of directly combining Eqs. \ref{eq:drdt} and
\ref{eq:hydrodrdt}, we apply the approach of de Ruijter \textit{et
al.} \cite{RCO:Lan:99,RCV:Lan:00} in order to make a direct comparison
with the combined model presented below. Using the same cylindrical
disk model, they neglect the flow perpendicular to the surface and
specify that the velocity at the upper edge of the cylinder is the
actual droplet spreading rate, $dR/dt$. With this approach, they
find that the hydrodynamic dissipation term can be written as

\begin{equation}
T\sum_{w}=6\pi R(t)\eta\phi[\theta(t)]\left(\frac{dR}{dt}\right)^{2}\ln\left[R(t)/a\right]\label{eq:dishydro}\end{equation}
 where $\phi\left(\theta\right)$is a geometric factor defined as\begin{equation}
\phi\left(\theta\right)=\frac{\sin^{3}\theta}{2-3\cos\theta+\cos^{3}\theta}\label{eq:combphi}\end{equation}
 and $a$ is an adjustable parameter that represents the radius of
the core region of the droplet, where the radial flow is negligible.
For the hydrodynamic model, they obtain\begin{equation}
\frac{d\theta}{dt}=-\left(\frac{\pi}{3V}\right)^{1/3}\Omega\left(\theta\right)\frac{\gamma\left(\cos\theta_{0}-\cos\theta\right)}{6\eta\phi\left(\theta\right)\ln\left[R/a\right]}.\label{eq:hydrodqdt}\end{equation}

Both types of dissipation are present in the spreading droplet. The
hydrodynamic mechanism is expected to dominate at low velocities and
small contact angles while the kinetic mechanism is expected to dominate
at high velocities and large contact angles \cite{BG:ACI:92}. We
include in our comparison a model developed by de Ruijter \textit{et
al.} \cite{RCO:Lan:99,RCV:Lan:00} containing both kinetic and hydrodynamic
terms. In this model, the velocity of the contact line is written
as\begin{equation}
\frac{dR}{dt}=\frac{\gamma\left[\cos\theta_{0}-\cos\theta\right]}{\zeta_{0}+6\eta\phi\left(\theta\right)\ln\left[R/a\right]}.\label{eq:combdrdt}\end{equation}
 Combining this with Eq. \ref{eq:drdt} gives\begin{equation}
\frac{d\theta}{dt}=-\left(\frac{\pi}{3V}\right)^{1/3}\Omega\left(\theta\right)\frac{\gamma\left(\cos\theta_{0}-\cos\theta\right)}{\zeta_{0}+6\eta\phi\left(\theta\right)\ln\left[R/a\right]}.\label{eq:combdqdt}\end{equation}

\subsection{Analysis of Models}

Fitting simulation data to the models described above requires both
the liquid/vapor surface tension and the bulk viscosity of the polymer.
The surface tension, $\gamma$, is obtained by first constructing
a slab of the polymer melt containing $10,000$ chains of $N=10$,
$5000$ chains of $N=20$, or $5000$ chains of $N=40$ centered in
the simulation box such that there are two surfaces perpendicular
to the $z$ direction. The simulations are run at temperature $T=1.0$
and pressure $P\simeq0$ without tail corrections. We leave out tail
corrections to the pressure in order to match the system of the spreading
droplet. The simulations are run until the two liquid/vapor interfaces
are equilibrated, as determined by the density profiles across the
interfaces. From the equilibrium values of the pressure, parallel
and perpendicular to the interfaces, $\gamma$ can easily be determined
from \cite{NBB:JCP:88}\begin{equation}
\gamma=\frac{1}{2}\int_{0}^{L_{z}}\left[p_{\perp}\left(z\right)-p_{\parallel}\left(z\right)\right]dz.\label{eq:surftens}\end{equation}
 The values for the surface tension are summarized in Table \ref{cap:bulkprops}.
These values can be compared to $\gamma=0.08\:\varepsilon/\sigma^{2}$
for a system of monomers \cite{SGL:PRE:99}.

\begin{table}

\caption{\label{cap:bulkprops}Bulk properties of bead-spring chains obtained
from MD simulation for $T=\varepsilon/k_{B}$, $P\simeq0.$}

\begin{tabular}{|c|c|c|c|c|c|}
\hline 
N&
$\rho$ ($\sigma^{-3})$&
$\gamma$ ($\varepsilon/\sigma^{2})$&
$\eta$ (m/$\tau\sigma$)&
$10^{3}$D ($\sigma^{2}/\tau$)&
$\zeta_{R}$ ($\tau^{-1})$\tabularnewline
\hline
\hline 
10&
0.8691&
0.85$\pm$0.02&
11.1$\pm$0.4&
6.17$\pm$0.06&
16.2\tabularnewline
\hline 
20&
0.8803&
0.92$\pm$0.02&
17.4$\pm$0.7&
3.04$\pm$0.03&
16.4\tabularnewline
\hline 
40&
0.8856&
0.95$\pm$0.02&
41.7$\pm$1.4&
1.23$\pm$0.01&
20.4\tabularnewline
\hline
\end{tabular}
\end{table}

The viscosity is computed from the equilibrium fluctuations of the
off-diagonal components of the pressure tensor \cite{AT:CSL:87}.
The pressure tensors are recorded from simulations of systems containing
melts of $500$ chains of the $N=10$ polymer, $250$ chains of the
$N=20$ polymer, and $500$ chains of the $N=40$ polymer at $T=1.0$
with the bulk pressure $P\simeq0$ without tail corrections. These
simulations are run at a timestep $\Delta t=0.006$ for up to $25,000\:\tau$.
The autocorrelation function of each off-diagonal component of the
stress tensor is calculated using the Numerical Recipes routine \textsc{correl}
\cite{PTV:NR:92}. The autocorrelation functions are averaged to improve
statistical uncertainty. From this, the viscosity can be calculated
using \cite{AT:CSL:87}\begin{equation}
\eta=\frac{V}{k_{B}T}\int_{0}^{\infty}dt\left\langle \sigma_{\alpha\beta}(t)\sigma_{\alpha\beta}(0)\right\rangle .\label{eq:visc}\end{equation}
 The results for $\eta$ are summarized in Table \ref{cap:bulkprops}.

Estimates of the friction coefficients, $\zeta_{R}$, obtained from
the melt simulations, are included in Table \ref{cap:bulkprops}.
The diffusion constant $D$ is determined from the mean square displacement
of the middle monomers of each chain and using the Rouse model one
can extract $\zeta_{R}$ from $D=\frac{k_{B}T}{mN\zeta_{R}}$ \cite{DE:TPD:86}. 

\begin{figure}
\includegraphics[width=8.5cm]{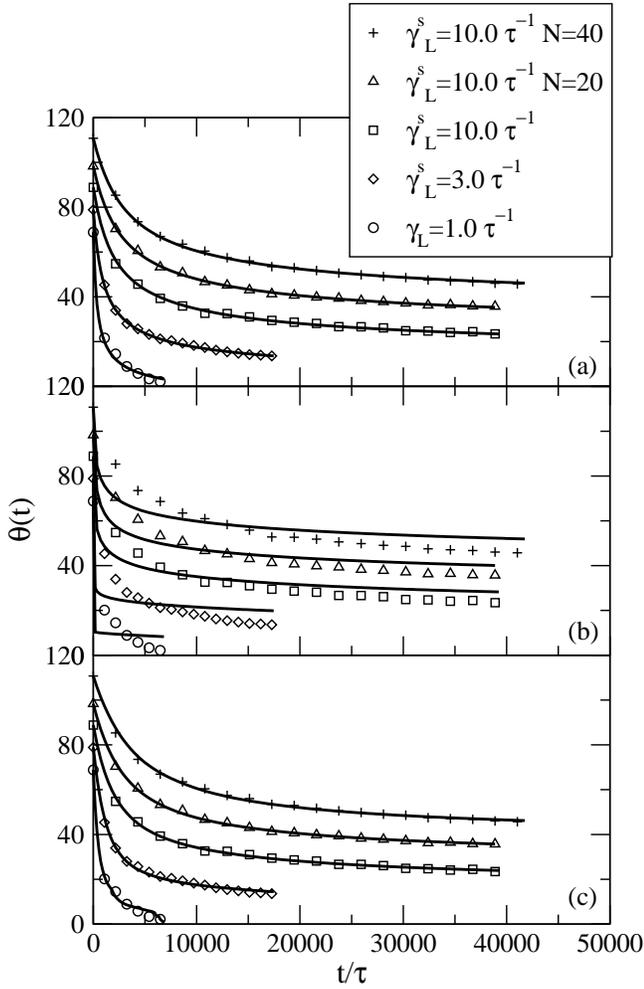}
\caption{\label{cap:flatfit}Fits to contact angle data (symbols) of (a) kinetic,
(b) hydrodynamic and (c) combined models for wetting droplets on a
flat surface with $\varepsilon_{w}=2.0$$\:\varepsilon$. The chain
length is $N=10$ unless otherwise specified. The Langevin thermostat
is applied to all monomers ($\gamma_{L}$) or just monomers near the
surface ($\gamma_{L}^{s}$). The data sets are shifted by $10^{o}$
increments (except for $\gamma_{L}^{s}=1.0\:\tau^{-1}$) for clarity. }
\end{figure}
\begin{figure}
\includegraphics[width=8.5cm]{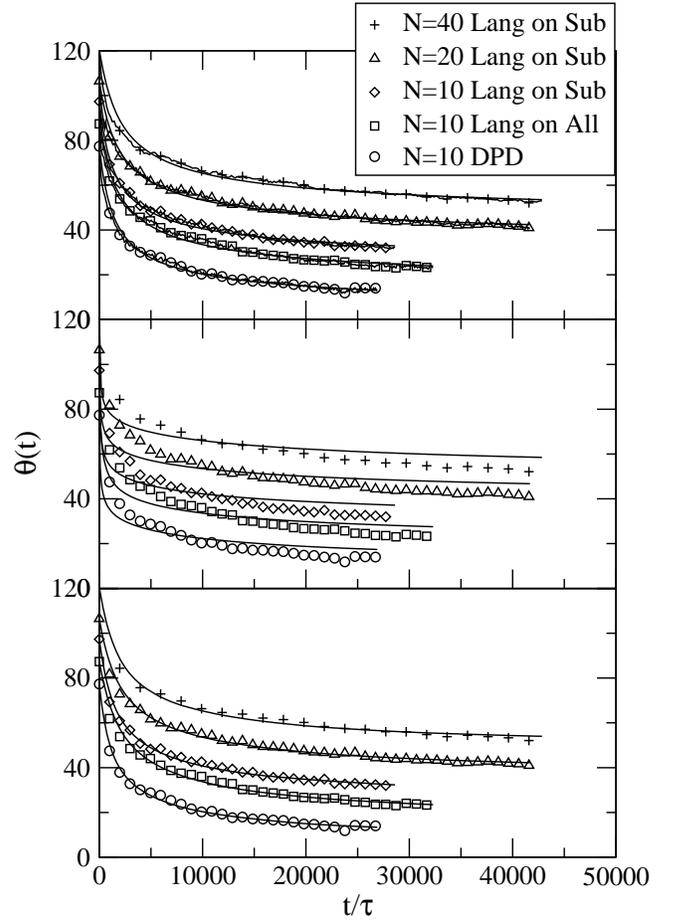}
\caption{\label{cap:subfit}Fits to contact angle data (symbols) of (a) kinetic,
(b) hydrodynamic and (c) combined models for wetting droplets on the
medium atomistic substrate with $\varepsilon_{12}=1.5$$\:\varepsilon$.
The Langevin thermostat is applied to all monomers or to just substrate
monomers. $\gamma_{L}=0.1\:\tau^{-1}$ for all cases except DPD where
$\gamma_{DPD}=0.1\:\tau^{-1}$. The data sets are shifted by $10^{o}$
increments (except for $N=10$ DPD) for clarity. }
\end{figure}
With the above values for the surface tension and viscosity, the simulation
data is fit to each of the models described above. The fit is performed
by taking initial guess values for the independent parameters and
integrating the expression for $d\theta/dt$ defined in one of the
equations \ref{eq:kindqdt}, \ref{eq:hydrodqdt}, or \ref{eq:combdqdt}.
The integration uses the fourth-order Runge-Kutta method to generate
a set of data, $\theta_{calc}(t)$. The parameters are varied using
the downhill simplex method \cite{PTV:NR:92} until the difference
between the model and simulation data, $\left|\theta_{calc}(t)-\theta(t)\right|/\theta(t)$,
is minimized.

The kinetic, hydrodynamic and combined models are fit to the contact
angles of droplets spreading on a flat surface in Fig. \ref{cap:flatfit}.
The Langevin thermostat is applied either to all monomers or only
to those near the surface. We find that both the kinetic and combined
models fit the data well despite the fact that they predict that the
friction coefficient, $\zeta_{0}$, is larger in the combined model
than in the kinetic model. The hydrodynamic model produces a very
poor fit to each data set as shown in Fig. \ref{cap:flatfit}b. The
best fit parameters for these models applied to data for wetting droplets
on a flat surface are shown in Table \ref{cap:modelparmsflat}. The
error reported for each model is calculated as \begin{equation}
\chi^{2}=\frac{1}{{\cal N}}\sum_{i=1}^{{\cal N}}\frac{\left|\theta_{calc}(t)-\theta(t)\right|^{2}}{\theta(t)}\label{eq:error}\end{equation}
 where ${\cal N}$ is the number of data points in each set of data.

\begin{table*}
\caption{\label{cap:modelparmsflat}Model parameters and error estimates resulting
from fits to contact angle data from simulations of wetting droplets
on a flat surface. Values for $\gamma_{L}$ and $\gamma_{L}^{s}$
are listed in the first column. $\varepsilon_{w}=2.0\:\varepsilon$.}

\begin{tabular}{|c|c|c|c|c|c|c|c|c|}
\hline 
&
&
Kinetic&
Hydrodynamic&
\multicolumn{2}{c|}{Combined}&
&
&
\tabularnewline
Thermostat&
N&
$\zeta_{0}\:\left(\frac{m}{\tau\sigma}\right)$&
a$\left(\sigma\right)$&
$\zeta_{0}\:\left(\frac{m}{\tau\sigma}\right)$&
a$\left(\sigma\right)$&
$\chi_{kin}^{2}$&
$\chi_{hydro}^{2}$&
$\chi_{comb}^{2}$\tabularnewline
\hline
$\gamma_{L}=1.0\:\tau^{-1}$&
10&
9.55&
44.40&
35.41&
71.23&
0.0022&
0.047&
0.0039\tabularnewline
\hline
$\gamma_{L}^{s}=3.0\:\tau^{-1}$&
10&
25.97&
42.29&
57.27&
71.55&
0.00028&
0.025&
0.0011\tabularnewline
\hline
$\gamma_{L}^{s}=10.0\:\tau^{-1}$&
10&
56.30&
38.14&
89.98&
83.80&
0.00024&
0.018&
0.00028\tabularnewline
\hline
$\gamma_{L}^{s}=10.0\:\tau^{-1}$&
20&
81.37&
38.83&
137.99&
84.77&
0.00015&
0.015&
0.00023\tabularnewline
\hline
$\gamma_{L}^{s}=10.0\:\tau^{-1}$&
40&
101.29&
38.63&
200.45&
86.49&
0.00022&
0.024&
0.00036\tabularnewline
\hline
\end{tabular}
\end{table*}

Figure \ref{cap:subfit} shows the kinetic, hydrodynamic, and combined
model fits to the contact angle data for wetting droplets on the medium
substrate. Again, the hydrodynamic model gives a significantly worse
fit to the data. The best fit parameters for these models for wetting
droplets on an atomistic substrate are shown in Table \ref{cap:modelparmssub}.
The kinetic and combined models give friction coefficients that are
generally larger than the bulk viscosity for the range of coupling
parameters used here. We find that, in contrast with previous work
by De Ruijter \textit{et al.} \cite{RCO:Lan:99,RCV:Lan:00}, the combined
model predicts a larger friction coefficient than the kinetic model.
Also, the hydrodynamic and combined models give a value of $a$ that
is on the order of the radius of the droplet, indicating that hydrodynamic
flow is not a dominant feature of the spreading of these droplets,
at least for the time scales accessible to simulation.

\begin{table*}
\caption{\label{cap:modelparmssub}}
Model parameters and error estimates resulting from fits to contact
angle data from simulations of wetting droplets on an atomistic substrate.
$\varepsilon_{12}=1.5$$\:\varepsilon$, $\gamma_{DPD}=\gamma_{L}=0.1$$\:\tau^{-1}$.
\begin{tabular}{|c|c|c|c|c|c|c|c|c|}
\hline 
&
&
Kinetic&
Hydrodynamic&
\multicolumn{2}{c|}{Combined}&
&
&
\tabularnewline
Thermostat&
N&
$\zeta_{0}\:\left(\frac{m}{\tau\sigma}\right)$&
a$\left(\sigma\right)$&
$\zeta_{0}\:\left(\frac{m}{\tau\sigma}\right)$&
a$\left(\sigma\right)$&
$\chi_{kin}^{2}$&
$\chi_{hydro}^{2}$&
$\chi_{comb}^{2}$\tabularnewline
\hline
DPD&
10&
36.7&
42.1&
53.4&
65.5&
0.001&
0.015&
0.001\tabularnewline
\hline
Lang on All&
10&
50.8&
38.1&
81.8&
70.0&
0.001&
0.015&
0.001\tabularnewline
\hline
Lang on Sub&
10&
38.0&
41.8&
64.9&
69.6&
0.001&
0.020&
0.001\tabularnewline
\hline
Lang on Sub&
20&
54.4&
43.2&
91.9&
68.3&
0.001&
0.020&
0.001\tabularnewline
\hline
Lang on Sub&
40&
65.5&
42.9&
126&
64.0&
0.002&
0.019&
 0.002\tabularnewline
\hline
\end{tabular}
\end{table*}

\section{\label{sec:Conclusions}Conclusions}

In this study, we perform molecular dynamics simulations of polymer
droplets that are roughly an order of magnitude greater in size than
those previously studied. We find this to be necessary to adequately
model the behavior of the precursor foot and the bulk material simultaneously.
Starting from a hemispherical droplet, we find that the precursor
foot forms immediately and spreads diffusively for each system where
the surface interaction strength is above the wetting/non-wetting
transition. The bulk region of the droplet spreads at a significantly
slower rate, but the data is too imprecise to distinguish between,
for example, a $t^{1/7}$or a $t^{1/10}$ scaling.

We perform spreading simulations on both an atomistically realistic
substrate and a perfectly flat surface. The simulations using a flat
surface exhibit the same behavior as the realistic substrate and greatly
improve the computational efficiency since the number of monomers
on the realistic substrate is typically several times greater than
the number of monomers in the droplet. However, to do so, it is critical
to apply a thermostat that couples only to monomers near the surface.
On an atomistic substrate, the most efficient method is to couple
only the substrate particles to the thermostat. This is computationally
faster than coupling all monomers to the DPD thermostat and leads
to the same results.

Several droplet spreading models have been developed to fit contact
angle data. A simple kinetic mechanism for energy dissipation fits
the data well and provides reasonable values for the friction coefficients,
which we verified through separate polymer melt simulations. Using
a combined model that adds a hydrodynamic energy dissipation mechanism
slightly improves the fit, but resulted in less accurate estimates
of the friction coefficients. The fact that we do not observe evidence
of hydrodynamic flow behavior may be due to the small droplet sizes
accessible to molecular dynamics simulation. Evidence for hydrodynamic
effects on spreading has been observed experimentally for macroscopic
drops \cite{T:JPD:79,PSS:JCI:01,GBS:LAN:97}. The length scale where
hydrodynamic effects become important remains an open question.

Future work will include studying the spreading behavior of binary
droplets and developing more realistic surface interactions.

\section{Acknowledgements}

We thank M. O. Robbins for helpful discussions. Sandia is a multiprogram
laboratory operated by Sandia Corporation, a Lockheed Martin Company,
for the United States Department of Energy's National Nuclear Security
Administration under Contract No. DE-AC04-94AL85000.

\newpage

\bibliographystyle{apsrev}
\bibliography{drop}

\end{document}